\documentclass[12pt]{article}
\usepackage[utf8]{inputenc}
\usepackage[T1]{fontenc}
\usepackage[english]{babel}
\usepackage{ifpdf} 
\usepackage{array,graphicx,abstract,hanging,fancyhdr}

\usepackage[margin=1in]{geometry}
\usepackage{times}
\usepackage{amsfonts}
\usepackage{amsmath}
\usepackage{amssymb}
\usepackage{bbold}
\usepackage{bm}
\usepackage{bbm}
\usepackage{natbib}
\usepackage{graphicx}
\usepackage{float}

\begin{document}

\title{Modelling for Poisson process intensities over irregular spatial domains}
\author{Chunyi Zhao and Athanasios Kottas \\
Department of Statistics, University of California, Santa Cruz}


\maketitle

\begin{abstract}
\noindent
We develop nonparametric Bayesian modelling approaches for Poisson processes, 
using weighted combinations of structured beta densities to represent the point 
process intensity function. For a regular spatial domain, such as the unit square, 
the model construction implies a Bernstein-Dirichlet prior for the Poisson process 
density, which supports general inference for point process functionals. 
The key contribution of the methodology is two classes of flexible and computationally 
efficient models for spatial Poisson process intensities over irregular domains. 
We address the choice or estimation of the number of beta basis densities, 
and develop methods for prior specification and posterior simulation for full inference 
about functionals of the point process. 
The methodology is illustrated with both synthetic and real data sets.
\end{abstract}

\noindent
KEY WORDS: Bayesian nonparametrics; Bernstein-Dirichlet process prior;
Markov chain Monte Carlo; Non-homogeneous Poisson process.

%
%

\section{Introduction}
\label{intro}

There has been an increasing interest in extracting information from locations in spatial data. 
For spatial point patterns, both the number and the locations of points are random.
Point pattern data is modelled as a realization, within compact domain $\mathcal{D}$, of a 
point process whose finite dimensional distribution defines the stochastic mechanism for 
the number and locations of the points. Independent increments along with a Poisson 
distributional assumption define the Poisson process.
A homogeneous Poisson process is equivalent to complete spatial randomness, that is, 
the point pattern generated is independently and identically uniformly distributed over 
$\mathcal{D}$. The practically relevant version is the non-homogeneous Poisson 
process (NHPP), which allows the point process intensity to differ by location. 
The NHPP is characterized by a non-negative, locally integrable intensity function 
$\lambda(s)$, such that: for any bounded subset $\mathcal{B}$ of the domain, 
the number of points in $\mathcal{B}$, $N(\mathcal{B})$, is $\text{Poisson}(\int_{\mathcal{B}}\lambda(s)\text{d}s)$ 
distributed; and, given $N(\mathcal{B})$, the point locations within $\mathcal{B}$ are independent
and identically distributed with density $\lambda(s)/\int_{\mathcal{B}}\lambda(u) \text{d}u$. 
Therefore, the NHPP likelihood corresponding to point pattern $\{ s_{1},...,s_{n} \}$, observed
in compact domain $\mathcal{D}$, can be expressed as: 
\begin{align}
p( \{ s_{1},...,s_{n} \} ; \lambda(s)) & \propto 
\exp\left( -\int_\mathcal{D} \lambda(s) \, \text{d}s \right)
\, \prod_{i=1}^n \lambda(s_i) 
\label{eq:poisonlik}
\end{align}
where $n \equiv N(\mathcal{D})$.
We consider the more common settings where $\mathcal{D} \subset \mathbb{R}$ or 
$\mathcal{D} \subset \mathbb{R}^2$. We place particular emphasis on spatial NHPPs, 
and more specifically on building flexible, computationally tractable 
models for spatial intensities defined over domains with irregular shapes.

Theoretical study of NHPPs can be found in \citet{Cressie1993-na} and \citet{Daley2008-up}, 
among other references. \citet{Diggle2003-tf} provides background on likelihood and classical 
nonparametric inference for spatial NHPPs. 
\citet{Moller2003-qd} discuss simulation-based inference for point processes. 
Regarding model-based methods for NHPPs, \citet{Gelfand2018-pz} categorize 
the main approaches in two general directions: modelling the trend surface for the intensity 
function $\lambda(s)$; and, factorizing the intensity function into the total intensity, 
$\Lambda =$ $\int_\mathcal{D} \lambda(s) \text{d}s$, and the NHPP density 
$f(s) =$ $\lambda(s)/\Lambda$, and modelling each separately.

The early Bayesian nonparametric approaches fall under the first category, focusing on modelling 
temporal NHPP cumulative intensity functions, $\int_{0}^t\lambda(s) \text{d}s$, with gamma, beta 
or general L\'evy process priors 
\citep{Lo1982,Lo:1992}. 
The next stage in this line of research involves mixture models for NHPP intensities built from 
non-negative kernels convolved with weighted gamma processes  
\citep{Lo1989-yq,WI1998,Ishwaran2004-zh,KNWJ2014}. 
Also in this direction are modelling approaches based on log-Gaussian Cox processes 
\citep{Moller1998-mn} under which the logarithm of the intensity function
is a realization of a Gaussian process. 
\citet{Adams2009-gb} proposed a related approach based on a logistic instead of 
logarithmic transformation to link the Gaussian process with the model for the 
intensity function. Modelling directly the intensity function $\lambda(s)$ brings 
computational challenges for full posterior inference due to the likelihood normalizing term, 
$\exp(-\int_\mathcal{D} \lambda(s) \, \text{d}s)$, especially under methods based on 
Gaussian process priors. Such challenges have been addressed through approximations of the stochastic integral \citep{Brix2001-dy,Brix2001-ba}, data augmentation \citep{Adams2009-gb},
and discretization of the observation domain $\mathcal{D}$ \citep{ISR_2012}.

Under the second direction, \cite{AK2006} and \cite{Kottas2007-xw} proposed an approach that 
connects the NHPP intensity function with the density function supported on the observation 
domain, and models the NHPP density with Dirichlet process mixture priors for density estimation. 
\citet{Taddy2012-fi} extend this modelling approach to marked Poisson processes, and 
\citet{Taddy2010}, \citet{KBMPO2012}, \citet{XKS2015} and \citet{RWK2017} develop 
hierarchical and dynamic models for NHPPs in the context of specific applications.
This modelling approach enables an inference framework that builds from well established 
methods for Dirichlet process mixtures, avoiding the computational challenges due to the 
NHPP likelihood normalizing component. However, it relies on a potentially restrictive
prior structure that models separately the NHPP density and the total intensity over 
the observation domain.

Inference methods for irregular domain spatial point process intensities have received 
limited attention in the Bayesian nonparametrics literature. We are only aware of the 
log-Gaussian Cox process approach of \cite{SILSR_2016}. Here, the irregular domain adds 
an extra level of complexity, which has been handled with an approximation to the Gaussian 
random field, an associated approximation to the NHPP likelihood, and using integrated 
nested Laplace approximation for fast, but approximate Bayesian inference.

Our main contribution is flexible modelling and computationally efficient inference 
for NHPPs over spatial domains with irregular shapes. The proposed models do not rely
on approximations of the NHPP likelihood and they can be efficiently implemented
with standard Markov chain Monte Carlo algorithms for full Bayesian inference
and uncertainty quantification. Moreover, in the
context of the more commonly studied setting of spatial NHPPs over regular domains, 
our modelling approach overcomes some of the limitations of existing Bayesian methods, 
while retaining the feature of flexible inference for general intensity shapes. 
%
%

We build the model for the NHPP intensity function from weighted combinations of 
Bernstein polynomial basis functions, that is, beta densities with specified shape 
parameters. Such parsimonious mixture representation is the key to achieve 
computationally tractable inference. In Section \ref{spatial}, we explore two 
modelling approaches for spatial Poisson process intensities over irregular domain,
taken without loss of generality to be a subset of the unit square. 
Under the first approach, the representation for the NHPP intensity is motivated 
by truncating over the irregular domain a NHPP density defined as a weighted 
combination of Bernstein densities on the unit square. The second approach targets 
directly the NHPP intensity modelling it as a structured weighted combination 
of truncated Bernstein densities. The two models offer different benefits while 
sharing the feature that the total intensity, $\Lambda$, can be readily expressed 
in terms of model parameters. Thus, both models bypass the challenge 
brought about from the NHPP likelihood normalizing term without separating 
the total intensity and NHPP density in the prior specification. In the case of 
regular domain, say the unit square, the two modelling approaches yield the same 
form for the NHPP intensity which implies a Bernstein-Dirichlet prior for the 
corresponding NHPP density. To highlight this connection and its implications 
in posterior simulation, we begin in the next section with the methodology for 
the simpler setting of temporal NHPPs.
%
%

%
%

\section{Methodology for temporal Poisson processes}
\label{temporal}

\subsection{Model formulation}
\label{temporal_model}

Here, we focus on modelling one-dimensional NHPPs observed
over a bounded domain, taken without loss of
generality to be the unit interval.
Motivated by Bernstein polynomial priors for densities with bounded
support, our model for the intensity function $\lambda(s)$ 
implies a Bernstein-Dirichlet process prior for the NHPP density, 
$f(s)=$ $\lambda(s)/\int_{0}^{1} \lambda(u) \text{d}u$,
for $s \in [0,1]$.

The Bernstein polynomial prior model for density $f$ on $[0, 1]$ is given by 
$f_{K}(s \mid F) = \sum_{k=1}^{K} \omega_{k}$\\$ \text{be}(s \mid k,K-k+1)$,
where $\text{be}(\cdot \mid a,b)$ is the beta density with mean $a/(a+b)$.
The mixture weights are defined through increments of a distribution 
function $F$ with support on $[0,1]$, such that $\omega_{k}=$ 
$F(k/K) - F((k-1)/K)$, for $k=1, \ldots, K$.
A distribution $F$ with flexible shape implies mixture weights that select 
the appropriate beta basis densities to achieve general shapes for density $f$. 
This motivates assigning a nonparametric prior to $F$, such as the 
Dirichlet process prior \citep{Ferguson:1973} which results in the
Bernstein-Dirichlet prior for density $f$ \cite[][]{Petrone1999-ne,Petrone1999-zy}. 
Theoretical support for the Bernstein polynomial model is provided by the fact 
that, as $K \rightarrow \infty$, $f_{K}(s \mid F)$ converges uniformly to the 
density of $F$ \citep{levasseur}; this result is also key to establishing 
Kullback-Leibler support and posterior consistency of the
Bernstein-Dirichlet prior for density estimation \citep{Petrone2002-lg}. 
Extensions of Bernstein polynomial prior models include density estimation 
on higher dimensional spaces \citep{Zheng2010-bw,BJQ_2015} and 
density regression \citep{BJQ_2017}.

Our modelling approach is motivated by the structure of the
distribution for the mixture weights, $(\omega_{1},...,\omega_{K})$, 
implied by a Dirichlet process prior, $\text{DP}(\alpha,F_{0})$,
on $F$, where $\alpha$ is the Dirichlet process precision parameter,
and $F_{0}$ the centering distribution with support on $[0,1]$. 
Based on the Dirichlet process definition, $(\omega_{1},...,\omega_{K})$, 
given $\alpha$, 
$F_{0}$, and $K$, follows a $\text{Dirichlet}(\alpha A_{1},...,\alpha A_{K})$ 
prior distribution, where $A_{k}=$ $F_{0}(k/K) - F_{0}((k-1)/K)$, for $k=1,...,K$. 
The key observation for the model is that the prior distribution for 
$(\omega_{1},...,\omega_{K})$ can be constructed through independent gamma
random variables. In particular, denoting by $\text{Ga}(a,b)$ the gamma distribution
with mean $a/b$, we have $\omega_{k}=$ $V_{k}/\{ \sum_{r=1}^{K} V_{r} \}$, where, 
for $k=1,...,K$, the $V_{k}$ are independently $\text{Ga}(\alpha A_{k},C)$ 
distributed, with $C > 0$ a constant.

The proposed model for one-dimensional NHPP intensities is given by:
\begin{equation}
\label{intensity-1d}
\begin{array}{c}
\lambda(s) \, = \, 
\sum\limits_{k=1}^{K} V_{k} \,\, \text{be}(s \mid k,K - k + 1),
\,\,\,\,\,\,\, s \in [0,1]  \\
V_{k}  \mid \alpha,F_{0} \, \stackrel{ind.}{\sim} \,
\text{Ga}(\alpha \{ F_0(k/K) - F_0((k-1)/K) \}, C),  \,\,\,\,\, k=1,...,K.
\end{array}
\end{equation}
The total intensity over the domain is
$\Lambda=$ $\int\nolimits_{0}^{1} \lambda(u) \text{d}u =$ $\sum_{k=1}^{K} V_{k}$,
and thus the NHPP density is given by 
$f(s)=$ $\lambda(s)/\{ \int\nolimits_{0}^{1} \lambda(u) \text{d}u \} =$
$\sum_{k=1}^{K} \omega_{k} \, \text{be}(s \mid k,K-k+1)$, where
$\omega_{k}=$ $V_{k}/\{ \sum_{r=1}^{K} V_{r} \}$. Hence, the implied 
model for the NHPP density is the Bernstein-Dirichlet prior model. 
Based on the Dirichlet process definition, this connection holds true for 
any $K$, that is, for any partition $\{ S_{k} = [(k-1)/K, k/K): k=1,...,K \}$ 
of the unit interval.

Note that, since $\Lambda=$ $\sum_{k=1}^{K} V_{k}$, we have 
$E(\Lambda \mid \alpha) =$ $\alpha/C$, which justifies using a general 
constant $C$ in the prior for the $V_{k}$, rather than taking $C=1$. 
That is, we wish to avoid the conflict of large values of $\alpha$ that 
would be needed under $C=1$ for large prior expected total intensity versus 
small values of $\alpha$ favoring non-standard intensity function shapes.

A $\text{Ga}(a_{\alpha}, b_{\alpha})$ prior is assigned to $\alpha$.
In terms of model economy, the uniform distribution is an appealing choice 
for $F_{0}$. This choice is sufficiently flexible in practice, as shown with 
the data examples of Section \ref{1d_data_example}, and it also yields 
a form for the average intensity that facilitates prior specification. 
With $F_{0}$ uniform, the prior mean for the intensity is constant, 
given by $\text{E}(\alpha)/C$, and it does not depend on $K$. 
Details on the prior mean for the intensity function are provided 
in the Appendix.

To explore posterior simulation under model (\ref{intensity-1d}), we consider 
two equivalent hierarchical model formulations for the observed point pattern 
$\{ 0 < s_{1} < ... < s_{n} < 1 \}$. 
As discussed above, there is an one-to-one correspondence between parameter
vectors $(V_{1},...,V_{K})$ and $\{ \Lambda, (\omega_{1},...,\omega_{K}) \}$, 
where $\omega_{k} =$ $F(S_{k})$, for $k=1,...,K$.
The prior distribution for $(V_{1},...,V_{K})$ in (\ref{intensity-1d}) 
corresponds to a $\text{DP}(\alpha,F_{0})$ prior for $F$, and a 
$\text{Ga}(\alpha, C)$ prior for $\Lambda$. Moreover, the NHPP likelihood in 
(\ref{eq:poisonlik}) can be conveniently expressed in terms of either 
parameterization:
\[
\begin{array}{rcl}
\prod\limits_{k=1}^{K} e^{-V_{k}} \, \prod\limits_{i=1}^{n} \left\{ 
\sum\limits_{k=1}^{K} V_{k} \, \text{be}(s_{i} \mid k,K - k + 1) \right\}
&  = & e^{-\Lambda} \, \Lambda^{n} \,
\prod\limits_{i=1}^{n} \left\{ 
\sum\limits_{k=1}^{K} F(S_{k}) \, \text{be}(s_{i} \mid k,K - k + 1) \right\} .
\end{array}
\]

Working with fixed $K$, the intensity formulation involves parameters 
$\{ (V_1, \ldots, V_K), \alpha\}$. Here, we introduce discrete latent variables 
$\{\xi_i: i = 1, \ldots, n\}$ indicating basis configuration for each 
time event. In a Gibbs sampler setting, the posterior full conditional 
for each $\xi_i$ is a discrete distribution with support on 
$\{ 1, \ldots, K \}$. Most importantly, given $\{\xi_i: i = 1, \ldots, n\}$ 
and $\alpha$, each $V_{k}$ follows a gamma posterior full conditional 
distribution, independently of $\{ V_{r}: r \neq k \}$. 
Lastly, $\alpha$ can be sampled using a Metropolis-Hastings step.

Alternatively, the density formulation builds from parameters 
$\{\Lambda, F, \alpha, K \}$. In this case, we introduce continuous latent variables 
$\{ \theta_i: i=1,...,n \}$ to leverage the Dirichlet process mixture 
representation for the NHPP density function: 
\begin{equation}
\label{1d_dpmm_density}
\begin{array}{c}
f(s_{i}) \, \equiv \, f_{K}(s_i \mid F) \, = \,
\int \sum_{k = 1}^K \mathbb{1}_{[\frac{k-1}{K}, \frac{k}{K})}(\theta_i) 
\, \text{be}(s_i \mid k, K - k + 1) \, \text{d}F(\theta_i) .
\end{array}
\end{equation}
A practically important feature of this formulation is that 
the number of basis densities, $K$, can be estimated without resorting to 
trans-dimensional Markov chain Monte Carlo algorithms. Here, the dimension of 
the parameter space does not change with $K$ because the posterior distribution 
does not involve the weights $\omega_{k}$, but rather the random distribution $F$ 
whose increments define the mixture weights.
Posterior simulation proceeds by first sampling from the marginal posterior of 
$\{ (\theta_1, \ldots, \theta_n), \Lambda, \alpha, K\}$, using Markov chain Monte Carlo
methods for Dirichlet process mixtures \citep{Escobar1995-cf,Neal2000}.
We then sample $(\omega_1, \ldots, \omega_K)$, given 
$(\theta_1, \ldots, \theta_n), \alpha, K$, from the Dirichlet distribution 
implied by the Dirichlet process conditional posterior distribution 
for $F$, given $(\theta_1, \ldots, \theta_n)$ and $\alpha$.
Finally, posterior samples for the NHPP density and intensity can be readily 
obtained, using their expressions under model (\ref{intensity-1d}).
Full details for both posterior simulation algorithms are given 
in the Appendix.

\subsection{Prior specification}
\label{1d_priors}

The prior for $\alpha$ and the value for $C$ can be specified 
using prior guesses at the total intensity, $\hat{\Lambda}$, and an average 
intensity value, $\hat{\lambda}$, over the observation window. We select 
$b_{\alpha}$ to provide a wide range for $\alpha$, and using 
$\text{E}(\lambda(s)) =$ $\text{E}(\alpha)/C$, set 
$\text{E}(\alpha)=$ $a_{\alpha}/b_{\alpha}=$ $C \hat{\lambda}$.
The marginal prior for the total intensity is
$p(\Lambda)=$ $\int \text{Ga}(\Lambda \mid \alpha,C) \,
\text{Ga}(\alpha \mid b_{\alpha} C \hat{\lambda},b_{\alpha}) \, 
\text{d}\alpha$. We use this expression to specify $C$ such that the 
median of $p(\Lambda)$ is equal to $\hat{\Lambda}$.

Note the connection between $\alpha$ and $K$ in controlling the shape 
of prior realizations for the NHPP intensity: for fixed $\alpha$, increasing 
$K$ results in intensities with larger number of modes and more local features;
and, for fixed $K$, decreasing $\alpha$ favors more variability and more 
localized structure in the intensities. In practice, it may suffice 
to estimate only $\alpha$ keeping $K$ fixed at sufficiently large values. 
Note that the beta densities in model (\ref{intensity-1d}) play the role of 
basis functions rather than of kernel densities in finite mixture models. 
Also key is the Dirichlet process
underlying the prior for the weights $V_{k}$, which select the
subset of beta densities that contribute more to the intensity
representation. As illustrated with simulated data in Section \ref{1d_data_example},
the discrete nature of the Dirichlet process prior can effectively guard 
against over-fitting if one conservatively chooses a larger value for $K$ 
than may be necessary for a particular point pattern.

A possible approach to specify $K$ involves prior information on 
the peak of the intensity, $\hat{\lambda}_{\max}$, without necessarily knowing 
where in the observation window the peak occurs. The idea is to find $K$ 
such that $\hat{\lambda}_{\max}$ matches a percentile of the prior 
distribution of $b^{*} V_{\max}$, where $V_{\max}=$ $\max\{V_k : k=1,...,K \}$,
and $b^{*}$ is the modal value of the $\text{beta}(2,K-1)$ density, 
that is, the first member of the Bernstein polynomial basis with a 
unimodal density. Under the uniform $F_{0}$ distribution, the 
$V_{k}$ are independently and identically gamma distributed, and thus the 
prior distribution of $V_{\max}$ is analytically available given $\alpha$; 
the marginal prior for $V_{\max}$ can also be readily explored 
through simulation. Table \ref{tab:table1} provides an illustration, using 
the 90th percentile of the marginal prior distribution for $V_{\max}$, under 
a $\text{Ga}(2.53,0.1)$ prior for $\alpha$, and with 
values for the peak intensity that are relevant to one of the data 
examples of Section \ref{1d_data_example}. 

\begin{table}[t!]
\caption{Illustration of the prior specification strategy for $K$. 
$Q_{0.9}^{V_{\max}}$ denotes the 90th percentile of the marginal prior distribution 
for $\max\{V_k: k=1,...,K \}$, and $b^{*}$ the modal value of the $\text{beta}(2,K-1)$ 
basis density.} 
\label{tab:table1}
\begin{center}
 \begin{tabular}{||c | c c c||} 
 \hline
 $K$ & $Q_{0.9}^{V_{\max}}$ & $b^{*}$ & $b^{*} \times Q_{0.9}^{V_{\max}}$ \\ [0.5ex] 
 \hline
  20    & 232.34  & 7.56 & 1755.85 \\
 \hline
  30    & 208.18  & 11.23 & 2338.0  \\
 \hline
 50    & 181.36  & 18.58 & 3370.34 \\ 
 \hline
 100   & 167.38  & 36.97 & 6188.82 \\ 
 \hline
\end{tabular}
\end{center}
\end{table}

As discussed in Section \ref{temporal_model}, using the intensity formulation, 
with fixed $K$, allows for a particularly simple and efficient method to implement 
model (\ref{intensity-1d}). The more general version of the model with random $K$ 
can be implemented at the expense of somewhat more complex Markov chain Monte Carlo 
algorithms for Dirichlet process mixtures. A discrete uniform or a truncated Poisson
distribution with support on $[K_{\min}, K_{\max}]$ are possible priors for $K$.

\subsection{Synthetic data examples for the temporal NHPP model}
\label{1d_data_example}

We consider two synthetic data sets generated from NHPPs with bimodal intensities.
For the first example, the intensity is $\lambda(s)=$ $700 \, \text{be}(s \mid 3, 18)$ + 
$300 \, \text{be}(s \mid 13, 8)$; this can be viewed as a special case of model 
(\ref{intensity-1d}) with $K=20$, although our prior model does not allow for 
zero weights. The second data set is obtained by logit-transforming points 
generated from a weighted combination of normal densities, $\lambda(s)=$ 
$400 \, \text{N}(s \mid -2.2, 1.0)$ + $600 \, \text{N}(s \mid 0.3, 0.8)$.  
We take large sizes for the simulated point patterns -- $n=993$ for the 
first, and $n=1037$ for the second example -- to ensure a meaningful comparison 
of posterior estimates with the true intensities.

We follow the approach of Section \ref{1d_priors} to specify $C=0.023$ and a 
$\text{Ga}(2.53,0.1)$
prior for $\alpha$, using for both data examples $1000$ as the prior estimate for the 
total intensity, and $1100$ for the average intensity. For the first example, we take
$K=20$, as well as $K=40$ to study the implication of using 
a number of basis densities that is twice as large as what should suffice. For the 
second example, assume we are told that the peak of the intensity has a value around 
$2300$. Then, referring to Table \ref{tab:table1}, $K=30$ can be taken as the number 
of basis densities, or, more conservatively, as a lower bound. We consider again a 
larger value, $K=50$, to check sensitivity of posterior inference results. We 
also implemented the density formulation for the second example, with a uniform 
prior on $[20, 60]$ assigned to $K$.

\begin{figure}[t!]
  \centering
  \includegraphics[width=0.95\linewidth]{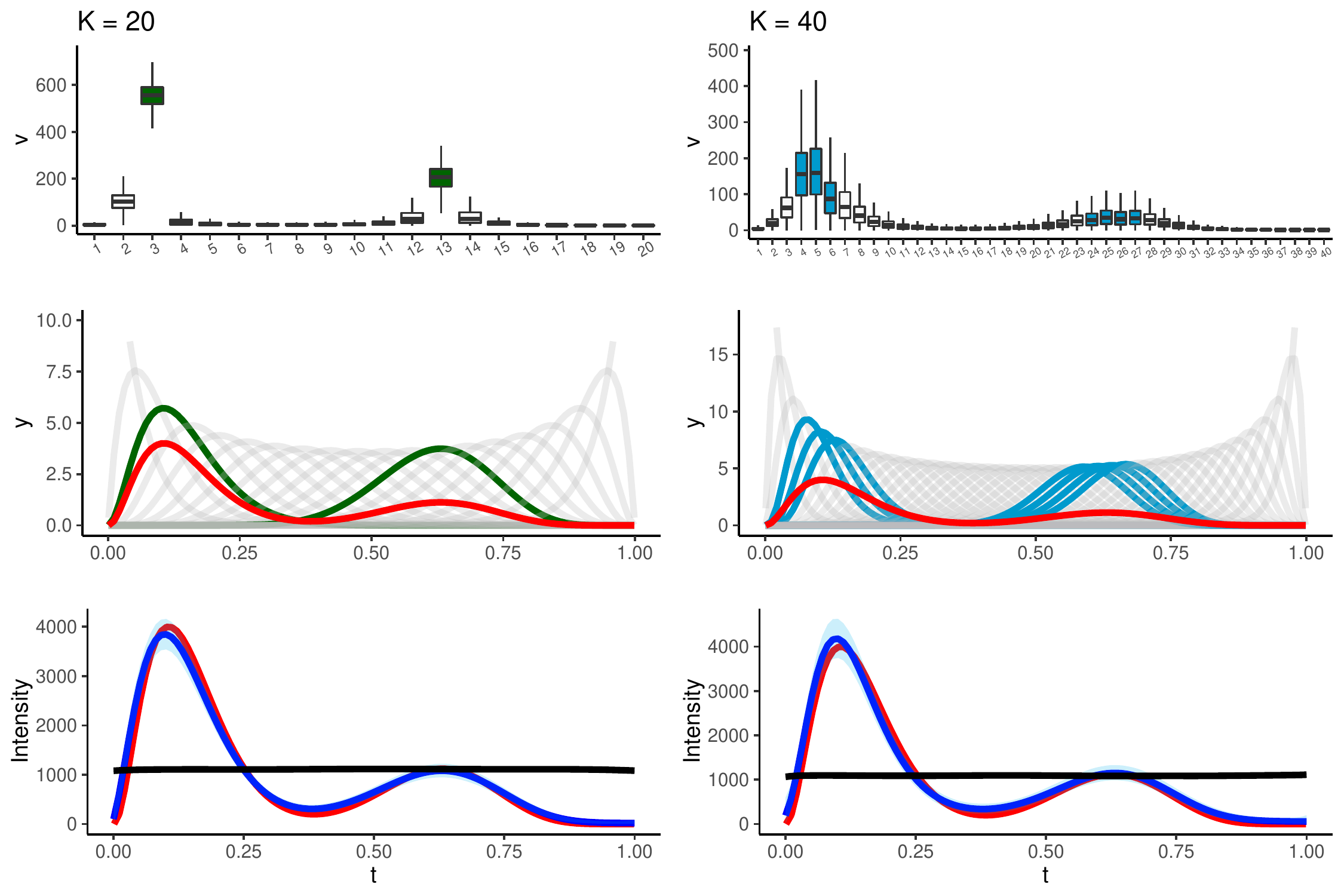}
\caption{
Beta mixture synthetic data example. Results under the 
intensity formulation with $K=20$ (left column) and $K=40$ (right column). 
Boxplots of posterior samples for the weights $V_k$ (first row), 
the beta basis densities corresponding to the largest $V_k$ (second row), 
and posterior mean (blue line) and 95\% interval estimates (light blue shaded bands) 
for the intensity function (third row). In the second and third rows, the red line
denotes the true density and intensity, respectively. In the third row, the 
black line indicates the prior mean for the intensity function.}
  \label{fig:Vk_boxplot}
\end{figure}

As shown in Fig. \ref{fig:Vk_boxplot}, the model is 
effective in estimating the weights that drive the bimodal intensity shape of the 
two-component beta mixture. Under $K=20$, it gives most weight to $V_3$ and 
$V_{13}$, that correspond to basis densities $\text{be}(s \mid 3, 18)$ and 
$\text{be}(s \mid 13, 8)$, whereas when $K = 40$, the model favors 6-7 basis 
densities with peaks in the same range as the two modes of the underlying intensity.
Hence, the model is able to achieve sparsity in estimation of the mixture 
weights when a surplus of basis densities are used, even though $F_{0}$ is 
a uniform distribution. Moreover, with the exception of some increase in the
width of posterior uncertainty bands, inference results for the intensity
function are similar under the two different choices for $K$.

\begin{figure}[t!]
  \centering
  \includegraphics[width=\linewidth]{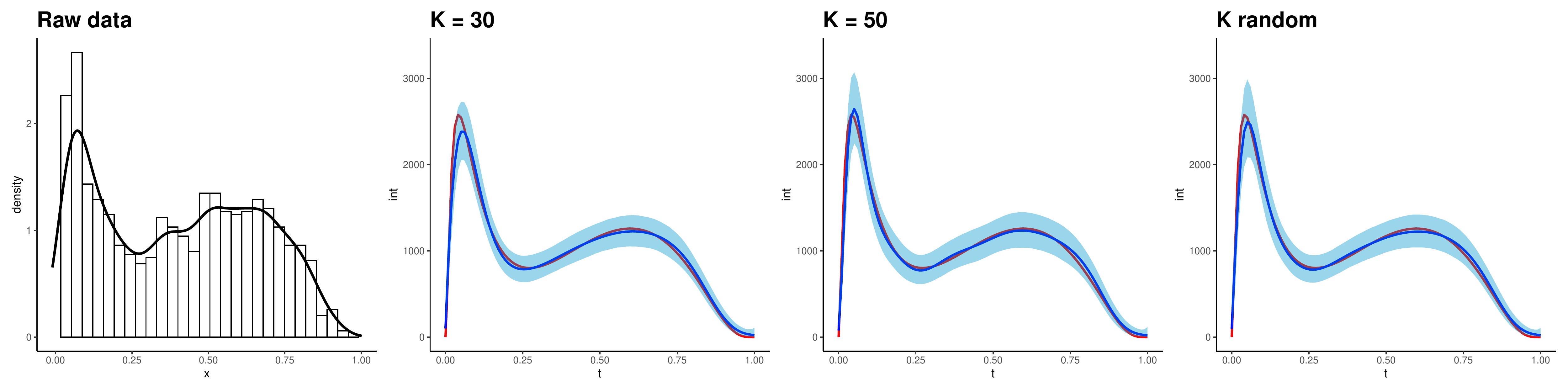}
  \caption{
Logit-normal mixture synthetic data example. From left to right, histogram of the 
simulated time points, and posterior mean (blue line) and 95\% interval 
estimates (light blue shaded bands) for the intensity function under $K=30$, $K=50$, and $K$ random. The red line in the last three panels
denotes the true intensity.}
  \label{fig:ln_int}
\end{figure}

This is also the case with the posterior inference results for the logit-normal mixture
data example; see Fig. \ref{fig:ln_int}. Under the density formulation, the 
posterior median for $K$ is $36$, with the 95\% credible interval given 
by $[22, 56]$. The intensity function under random $K$ has similar point estimate 
and a slightly tighter uncertainty band compared to that under $K = 50$.

%
%

\section{Modelling approaches for Spatial Poisson processes}
\label{spatial}

We begin with the case of a regular domain for the spatial NHPP, taken without 
loss of generality to be the unit square, such that $s \equiv$ $(x,y) \in [0, 1]^{2}$.
The extension of the Bernstein polynomial basis consists of products of beta
densities. More specifically, the basis density with index 
$(k_x, k_y)$, for $k_x, k_y = 1,...,K$, is defined as     
\begin{align}
\phi_{k_x, k_y}(x,y) &= \text{be}\left(x \mid k_x, K - k_x + 1\right) 
\text{be}\left(y \mid k_y, K - k_y +1\right), \,\,\,\,\,
(x,y) \in [0, 1]^2 .
\end{align}
Although the number of basis densities may be different in the 
$x$ and $y$ dimensions, we use the more parsimonious form with $K_{x}=K_{y}=K$.

Then, we can extend model (\ref{intensity-1d}) to the following model for 
spatial NHPP intensities over $[0, 1]^2$:
\begin{equation}
\label{intensity-2d}
\begin{array}{c}
\lambda(x,y) \, = \,
\sum\limits_{k_x,k_y = 1}^{K} V_{k_x, k_y} \,
\phi_{k_x, k_y}(x,y),  \,\,\,\,\,\,\,  (x,y) \in [0, 1]^2 \\
V_{k_x, k_y}  \mid \alpha,F_{0} \, \stackrel{ind.}{\sim} \,
\text{Ga}(\alpha F_0(S_{k_x, k_y}), C),  \,\,\,\,\,  k_x, k_y=1,...,K
\end{array}
\end{equation}
where $S_{k_x, k_y}=$ $[(k_x-1)/K, k_x/K) \times [(k_y-1)/K, k_y/K)$, 
and $F_0(S_{k_x, k_y})$ is the probability of $S_{k_x, k_y}$ under a 
specified distribution $F_{0}$ on $[0,1]^{2}$; in particular, 
$F_0(S_{k_x, k_y})=$ $1/K^2$ under the uniform distribution for $F_{0}$.

Again, the total intensity over the domain is readily
obtained as $\Lambda=$ $\int_0^1\!\int_0^1 \lambda(x,y) \, \text{d}x \text{d}y =$
$\sum_{k_x, k_{y} = 1}^{K} V_{k_x, k_y}$, and the NHPP density is given by 
$f(x, y)=$ $\sum_{k_x, k_y = 1}^{K} \omega_{k_x, k_y} \, \phi_{k_x, k_y}(x,y)$,
where $\omega_{k_x, k_y}=$ $V_{k_x, k_y}/\{ \sum\nolimits_{k_x, k_y=1}^{K} V_{k_x, k_y} \}$.
The implied prior distribution for the mixture weights $\{ \omega_{k_x,k_y} \}$
corresponds to constructing them through $\omega_{k_x,k_y} =$ $F(S_{k_x, k_y})$,
where $F$ is a random distribution on $[0,1]^{2}$ assigned a $\text{DP}(\alpha, F_0)$ 
prior.

We thus retain the connection between the intensity prior model 
in (\ref{intensity-2d}) and the two-dimensional Bernstein-Dirichlet prior 
model for the NHPP density, as well as the equivalent hierarchical model 
formulations for the data. Again, the implied $\text{Ga}(\alpha, C)$ prior 
for $\Lambda$ ensures the coherence between the intensity and density prior 
models, the latter comprising parameters  $\{ \Lambda, F, \alpha, K\}$.
Extending the approaches outlined in Section \ref{temporal_model}, 
posterior simulation can be implemented using either the 
intensity or density formulation. The prior mean intensity is 
$\text{E}(\lambda(x,y))=$ $\text{E}(\alpha)/C$, and thus the
prior specification approach of Section \ref{1d_priors} can be 
extended to model (\ref{intensity-2d}).

To achieve our main objective of flexible inference for NHPP spatial intensities
recorded over irregular domain $\mathcal{D} \subset [0, 1]^2$, we propose two 
different modelling approaches. Under the first model, presented in 
Section \ref{section_int}, the intensity formulation is motivated by truncating 
over $\mathcal{D}$ the NHPP density $f(x,y)$ defined on $[0, 1]^2$.
The second model, developed in Section \ref{section_den}, builds the basis 
representation for the intensity through the corresponding density which is 
defined as a mixture of truncated beta densities over $\mathcal{D}$ with 
weights induced by a random distribution $F$ on $\mathcal{D}$.
In both cases, the Bernstein polynomial prior structure is especially 
attractive to model spatial point process intensities over irregular domains, 
a practically relevant problem that, arguably, has not been fully addressed 
in the Bayesian nonparametrics literature.
%
%

\subsection{The intensity model}
\label{section_int}

Under the first modelling perspective, the representation for the NHPP intensity $\lambda_{\mathcal{D}}(x, y)$ over irregular domain $\mathcal{D}$ is revealed by 
the expression for $f_{\mathcal{D}}(x, y)$, the NHPP density truncated on 
$\mathcal{D}$. In particular, 
\begin{equation}
\label{int_f_D}
f_{\mathcal{D}}(x, y) = 
\frac{f(x,y)}{\int\!\int_{\mathcal{D}} f(u,v) \, \text{d}u \text{d}v} = 
\sum_{k_x, k_y =1}^{K} \frac{V_{k_x, k_y} B_{k_x, k_y}}{
\sum\nolimits_{k_x, k_y =1}^{K} V_{k_x, k_y} B_{k_x, k_y}}
\, \phi^{*}_{k_x, k_y}(x,y), \,\,\, (x,y) \in \mathcal{D}
\end{equation}
where $B_{k_x, k_y} =$ 
$\int\!\int_{\mathcal{D}} \phi_{k_x, k_y}(x,y) \, \text{d}x \text{d}y$, 
$\phi^{*}_{k_x, k_y}(x,y) =$ $\phi_{k_x, k_y}(x,y)/B_{k_x, k_y}$ are the basis
densities truncated on $\mathcal{D}$, and 
we have used the fact that $\omega_{k_x, k_y} B_{k_x, k_y}/
\{ \sum\nolimits_{k_x, k_y =1}^{K} \omega_{k_x, k_y} B_{k_x, k_y} \} =$\\
$V_{k_x, k_y} B_{k_x, k_y}/
\{ \sum\nolimits_{k_x, k_y =1}^{K} V_{k_x, k_y} B_{k_x, k_y} \}$. 
The implied model for the intensity function is:
\begin{align}
\label{intensity-2d_irregular}
\lambda_{\mathcal{D}}(x, y) & = \sum_{k_x, k_y=1}^{K} 
V_{k_x, k_y} B_{k_x, k_y} \, \phi^{*}_{k_x, k_y}(x,y), 
\,\,\,\,\,\,\,  (x,y) \in \mathcal{D}
\end{align}
where $V_{k_x, k_y}  \mid \alpha \stackrel{ind.}{\sim}$
$\text{Ga}(\alpha/K^{2} , C)$, for $k_x, k_y=1,...,K$, taking the uniform 
distribution for $F_{0}$, and placing a $\text{Ga}(a_{\alpha},b_{\alpha})$
prior on $\alpha$.

Evidently, (\ref{intensity-2d}) and (\ref{intensity-2d_irregular}) agree 
when $\mathcal{D}$ is the unit square.
Note that $B_{k_x, k_y}$ will be small for basis densities with significant mass  
outside $\mathcal{D}$. Hence, although model (\ref{intensity-2d_irregular}) uses 
all $K^{2}$ basis densities, the constants $B_{k_x, k_y}$ provide an additional 
adjustment to the one applied by the random coefficients $V_{k_x, k_y}$.  
The overhead cost of computing the normalizing constants $B_{k_x, k_y}$ is very 
small, since, with fixed $K$, they need to be computed only once. 

For posterior simulation, we introduce a pair of latent variables, 
$(\xi_{i},\eta_{i})$, for each point in the spatial point pattern,
$\{ (x_{i},y_{i}): i=1,...,n \}$, to identify the corresponding basis 
density. Then, the hierarchical model for the data can be written as:
\begin{equation}
\label{irr_spatial_full} 
\begin{array}{rcl}
\{ (x_{i},y_{i}) \} \mid V, \{ (\xi_{i},\eta_{i}) \} & \sim &
\exp\left( - \sum\limits_{k_x,k_y = 1}^{K} V_{k_x, k_y} B_{k_x, k_y}  \right) \,
\prod\limits_{i=1}^{n} \Lambda_{\mathcal{D}} \, 
\phi^{*}_{\xi_{i}, \eta_{i}}(x_{i},y_{i}) 
\\
(\xi_{i},\eta_{i}) \mid V & \stackrel{i.i.d.}{\sim} & 
\sum\limits_{k_x, k_y =1}^{K} \frac{V_{k_x, k_y} B_{k_x, k_y}}{\Lambda_{\mathcal{D}}} \,
\delta_{(k_x, k_y)}(\xi_{i},\eta_{i}),  \quad i=1,...,n  
\end{array}
\end{equation}
where $V=$ $\{ V_{k_x, k_y}: k_x,k_y = 1,...,K \}$, and $\Lambda_{\mathcal{D}}$
is the total intensity over the irregular domain, $\Lambda_{\mathcal{D}} =$
$\int\!\int_{\mathcal{D}} \lambda_{\mathcal{D}}(x,y) \, \text{d}x \text{d}y =$
$\sum_{k_x, k_y=1}^{K} V_{k_x, k_y} B_{k_x, k_y}$.

As with models (\ref{intensity-1d}) and (\ref{intensity-2d}), the form of 
the NHPP likelihood normalizing term implied by the intensity model 
(\ref{intensity-2d_irregular}) results in efficient posterior simulation 
with remarkably simple updates for parameters $\{ V_{k_x, k_y} \}$; given the 
$(\xi_{i},\eta_{i})$ and $\alpha$, the $V_{k_x, k_y}$ are conditionally independent
and gamma distributed. The Markov chain Monte Carlo posterior simulation algorithm 
is detailed in the Appendix.
%
%

In contrast to models (\ref{intensity-1d}) and (\ref{intensity-2d}),
the NHPP density in (\ref{int_f_D}) does not follow the Bernstein-Dirichlet 
prior. Consequently, we do not have a Dirichlet process mixture representation 
for the hierarchical model for the data, which allows estimating
$K$ without trans-dimensional posterior simulation algorithms. Therefore, 
practical implementation of model (\ref{irr_spatial_full}) requires specifying $K$. 
In practice, we recommend sensitivity analysis for the value of $K$. With $K$
selected, the approach of Section \ref{1d_priors} 
can be used to specify the prior for $\alpha$ and the value for $C$. The prior 
mean of the intensity function is again given by $\text{E}(\lambda_{\mathcal{D}}(x, y))=$ 
$\text{E}(\alpha)/C$, and, although $\Lambda_{\mathcal{D}}$ no longer follows
a gamma prior distribution, given $\alpha$, its marginal prior can
be easily developed by simulation.

%
%

\subsection{The density model}
\label{section_den}

Here, we seek to develop a model for the irregular domain intensity that corresponds
to a Bernstein-Dirichlet prior for the associated density, in the spirit of models
(\ref{intensity-1d}) and (\ref{intensity-2d}).
To this end, we define directly the density $f_{\mathcal{D}}(x, y)$ as a mixture of 
truncated beta basis densities:
\begin{equation}
\label{2d-density-ireg_dens}
f_{\mathcal{D}}(x, y) \, = \,
\sum\limits_{(k_x, k_y) \in J_{K}} \omega_{k_x, k_y}^{*} \, \phi_{k_x, k_y}^{*}(x, y),
\,\,\,\,\,\,\, (x,y) \in \mathcal{D}
\end{equation}
where $J_{K} =$ $\{ (k_x, k_y): S_{k_x, k_y} \cap \mathcal{D} \neq \emptyset \}$ is the index 
set for all non-empty intersections, $S^{*}_{k_x, k_y} =$ $S_{k_x, k_y} \cap \mathcal{D}$,
of the unit square partitioning sets $\{ S_{k_x, k_y}: k_x,k_y = 1,...,K \}$ with $\mathcal{D}$.
The mixture weights are defined as $\omega_{k_x, k_y}^{*} = F(S_{k_x, k_y}^{*})$, where $F$ 
is a random distribution on $\mathcal{D}$ following a $\text{DP}(\alpha, F_0)$ prior, with
$F_{0}$ taken to be the uniform distribution on $\mathcal{D}$.

We now define the model for the irregular domain spatial intensity as
\begin{equation}
\label{2d-density-ireg}
\begin{array}{c}
\lambda_{\mathcal{D}}(x, y) \, = \,
\sum\limits_{(k_x,k_y) \in J_{K}} V^{*}_{k_x, k_y} \,
\phi^{*}_{k_x, k_y}(x,y),  \,\,\,\,\,\,\,  (x,y) \in \mathcal{D} \\
V^{*}_{k_x, k_y}  \mid \alpha \, \stackrel{ind.}{\sim} \,
\text{Ga}(\alpha F_0(S^{*}_{k_x, k_y}), C),  \,\,\,\,\,  (k_x, k_y) \in J_{K}
\end{array}
\end{equation}
such that the density $f_{\mathcal{D}}(x, y) =$
$\lambda_{\mathcal{D}}(x, y) / \{ \int\!\int_{\mathcal{D}} \lambda_{\mathcal{D}}(u, v)
\, \text{d}u \text{d}v \}$ follows the prior model in (\ref{2d-density-ireg_dens}).
Again, the key link between parameterizations $\{ V^{*}_{k_x, k_y}: (k_x, k_y) \in J_{K} \}$
and $\{ \Lambda_{\mathcal{D}}, \{ \omega^{*}_{k_x, k_y}: (k_x, k_y) \in J_{K} \} \}$
is the practical expression for the total intensity $\Lambda_{\mathcal{D}} =$
$\int\!\int_{\mathcal{D}} \lambda_{\mathcal{D}}(x,y) \, \text{d}x \text{d}y =$
$\sum_{(k_x, k_y) \in J_{K}} V^{*}_{k_x, k_y}$, and its $\text{Ga}(\alpha, C)$ prior 
implied by (\ref{2d-density-ireg}).

For a spatial point pattern $\{ (x_{i},y_{i}): i=1,...,n \}$ recorded over $\mathcal{D}$,
we can write the NHPP likelihood in terms of either the intensity of density formulation:
\[
\begin{array}{c}
\exp\left( - \sum\nolimits_{(k_x,k_y) \in J_{K}} V^{*}_{k_x, k_y} \right) \,
\prod\limits_{i=1}^{n} \left\{ 
\sum\nolimits_{(k_x,k_y) \in J_{K}} V^{*}_{k_x, k_y} \, \phi^{*}_{k_x, k_y}(x_{i},y_{i})
\right\} \\
= \, \exp(-\Lambda_{\mathcal{D}}) \, \Lambda_{\mathcal{D}}^{n} \,
\prod\limits_{i=1}^{n} \left\{ 
\sum\nolimits_{(k_x,k_y) \in J_{K}} F(S^{*}_{k_x, k_y}) \, \phi^{*}_{k_x, k_y}(x_{i},y_{i})
\right\} .
\end{array}
\]
To explore the posterior distribution for $\{ \Lambda_{\mathcal{D}}, F, \alpha, K \}$ under 
the density formulation, we introduce bivariate continuous latent variables $\{ z_i \}$ 
to write the hierarchical model for the data:
\begin{equation}
\label{irr_spatial_den_full} 
\begin{array}{rcl}
\{ (x_{i},y_{i}) \} \mid \{ z_i \}, \Lambda_{\mathcal{D}}, K  & \sim &  
\exp( - \Lambda_{\mathcal{D}} ) \, \Lambda_{\mathcal{D}}^{n} \,
\prod_{i=1}^{n} \left\{ 
\sum\nolimits_{(k_x,k_y) \in J_{K}}
\mathbb{1}_{S^{*}_{k_x, k_y}}(z_i) \, \phi^{*}_{k_x, k_y}(x_{i},y_{i}) 
\right\}
\\
z_i \mid F & \stackrel{i.i.d}{\sim} & F,  \,\,\,\,\, i = 1,\ldots, n  \\
F \mid \alpha & \sim & \text{DP}(\alpha, F_0)
\end{array}
\end{equation}
where $\Lambda_{\mathcal{D}} \mid \alpha \sim \text{Ga}(\alpha, C)$, with 
a $\text{Ga}(a_{\alpha},b_{\alpha})$ prior placed on $\alpha$, and with a discrete 
uniform or a truncated Poisson prior distribution for $K$ with support 
on $[K_{\min}, K_{\max}]$. The Markov chain Monte Carlo algorithm for model (\ref{irr_spatial_den_full}) is given in the Appendix.
The posterior simulation method is more involved than the one for the intensity
model of Section \ref{section_int}, but it allows for estimation of $K$ without 
trans-dimensional computational techniques.

The marginal prior for the total intensity is $p(\Lambda_{\mathcal{D}})=$ 
$\int \text{Ga}(\Lambda_{\mathcal{D}} \mid \alpha,C) \,
\text{Ga}(\alpha \mid a_{\alpha}, b_{\alpha}) \, 
\text{d}\alpha$. Under model (\ref{2d-density-ireg}), there is no closed-form 
expression for $\text{E}(\lambda_{\mathcal{D}}(x, y))$, but $\text{E}(\alpha)/C$
is an approximate lower bound for the prior mean intensity. With this caveat,
the approach of Section \ref{1d_priors} can be used to specify the prior 
hyperparameters for $\alpha$ and the value for $C$. The earlier approach to 
specify $K$ can be used here to guide the choice of the support for the prior on $K$.

\section{Synthetic data examples}
\label{sim_study}

\begin{figure}
  \centering
  \includegraphics[width=0.95\linewidth]{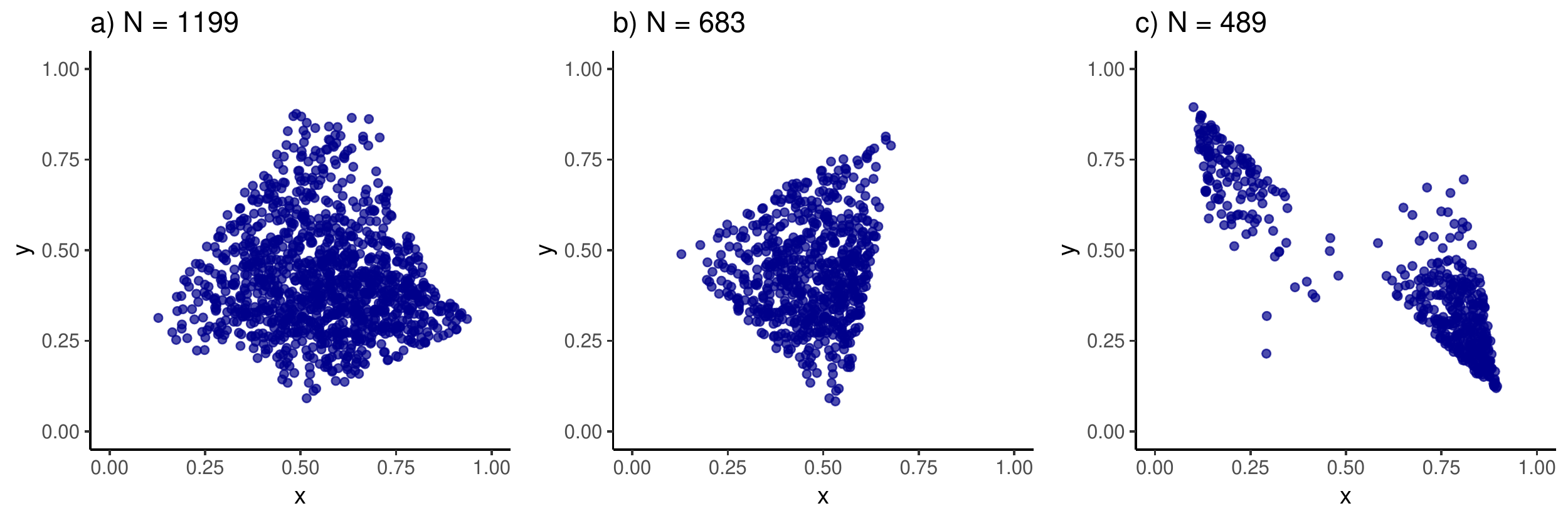}
\caption{Synthetic spatial point patterns for the irregular domain simulation study.
The size of each point pattern is shown in the corresponding panel.}
  \label{fig:2diregdata}
\end{figure}
\begin{figure}
  \centering
  \includegraphics[width=0.91\linewidth]{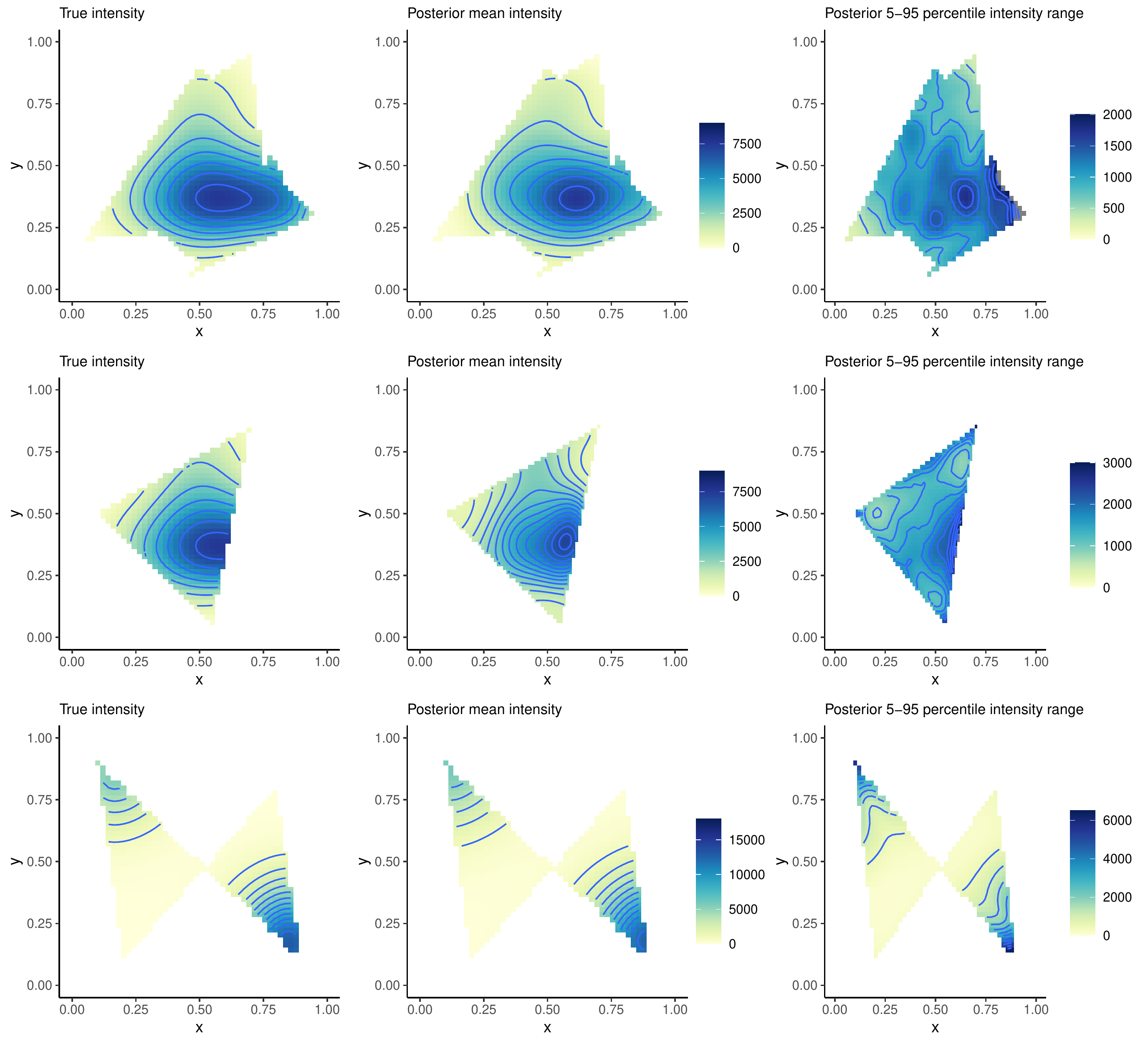}
\caption{Results for the data in Fig. \ref{fig:2diregdata} under the intensity model. The left 
panel shows the true intensity function, the middle panel the posterior mean intensity estimate, and 
the right panel a posterior uncertainty estimate in the form of the difference between the 95th and 
5th percentiles of the posterior distribution for the intensity function.}
  \label{fig:2diregb}
\end{figure}

We study inference results under both the intensity and density model,
using point patterns generated under 
three different scenarios for the irregular shape of the spatial NHPP. The 
synthetic point patterns are plotted in Fig. \ref{fig:2diregdata}, and the true intensities, 
as well as their corresponding polygonal domain, are shown in Fig. \ref{fig:2diregb}. 
For cases (a) and (b), the true NHPP density is a mixture of two bivariate logit-normal 
densities, truncated over the respective domain, which results in a unimodal intensity. 
Case (c) arises from truncating a mixture of bivariate beta densities that accumulates 
most of its mass at the $(0,1)$ and $(1,0)$ corners of the unit square. 

%
%

For all three cases, the intensity model (\ref{irr_spatial_full}) is implemented with 
$C = 0.05$, a $\text{Ga}(2, 0.01)$ prior for $\alpha$, and with $K = 20$.
The posterior mean and uncertainty estimates reported in Fig. \ref{fig:2diregb} demonstrate 
that the model recovers well the underlying intensity shapes over the different polygons.

We also applied the density model (\ref{irr_spatial_den_full}), using for all three data sets, 
a $\text{Ga}(5, 0.1)$ prior for $\alpha$, $C = 0.01$, and a discrete uniform prior on $[5, 25]$ 
for $K$. The posterior probability for $K$ at its posterior mode was: 
$\text{Pr}(K = 13 \mid \text{data}) = 0.89$ in case (a), 
$\text{Pr}(K = 12 \mid \text{data}) = 0.99$ in case (b), and
$\text{Pr}(K = 9 \mid \text{data}) = 0.81$ in case (c).
The posterior mean and uncertainty estimates under the density model were similar to the ones 
reported in Fig. \ref{fig:2diregb} under the intensity model.


As an additional illustration, we consider a point pattern of size $n = 303$ drawn from 
a NHPP with density $0.7 \, \text{be}(x \mid 4, 17) \text{be}(y \mid 10, 11)$ 
+ $0.3 \, \text{be}(x \mid 12, 9) \text{be}(y \mid 4, 17)$ truncated to the triangle with 
vertices $\{ (0.01, 0.01), (0.2, 0.9), (0.9, 0.1) \}$, and with total intensity $300$. 
Here, the truth is designed to resemble the intensity model with $K=20$, and we test the 
performance of the density model in estimating $K$ and other NHPP functionals.

\begin{figure}
  \centering
  \includegraphics[width=0.95\linewidth]{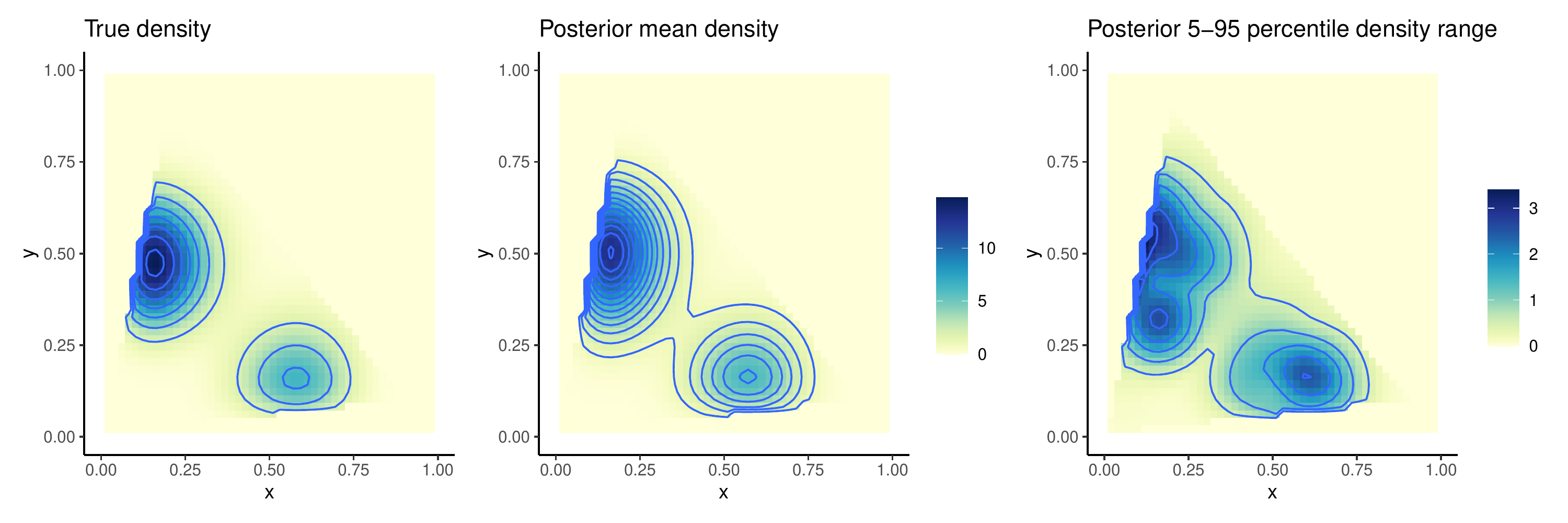}
  \caption{Results for the synthetic spatial point pattern generated from NHPP density 
$0.7 \, \text{be}(x \mid 4, 17) \text{be}(y \mid 10, 11)$ + 
$0.3 \, \text{be}(x \mid 12, 9) \text{be}(y \mid 4, 17)$ truncated to the triangle with 
vertices $\{ (0.01, 0.01), (0.2, 0.9), (0.9, 0.1) \}$. The left panel includes the 
true density. Based on the density model, the middle panel plots the posterior mean density 
estimate, and the right panel an uncertainty estimate given by the difference between the 
95th and 5th percentiles of the posterior distribution of the density function.}
  \label{fig:2d_ireg_den_model}
\end{figure}

Model (\ref{irr_spatial_den_full}) is implemented with a Ga$(2, 0.01)$ prior for $\alpha$, 
$C = 0.01$, and a discrete uniform prior for $K$ with support on $[15, 25]$. The posterior 
mean and uncertainty estimates in Fig. \ref{fig:2d_ireg_den_model} show that the underlying 
bimodal density shape is recovered well, taking into account the moderate size of the point 
pattern. The posterior mean for the total intensity is $301.1$, and the 95\% posterior credible 
interval is given by $(267.3, 334.3)$. The 95\% posterior credible interval for $K$ is 
$[19, 25]$, and the posterior mode is 20, with $\text{Pr}(K = 20 \mid \text{data}) = 0.46$.
We note that increasing the size of the simulated point pattern results in posterior 
distributions for $K$ that are more concentrated around $K=20$.

%
%

\section{Boston crime data analysis}
\label{boston}

\begin{figure}[t!]
\centering
\includegraphics[width=0.95\linewidth]{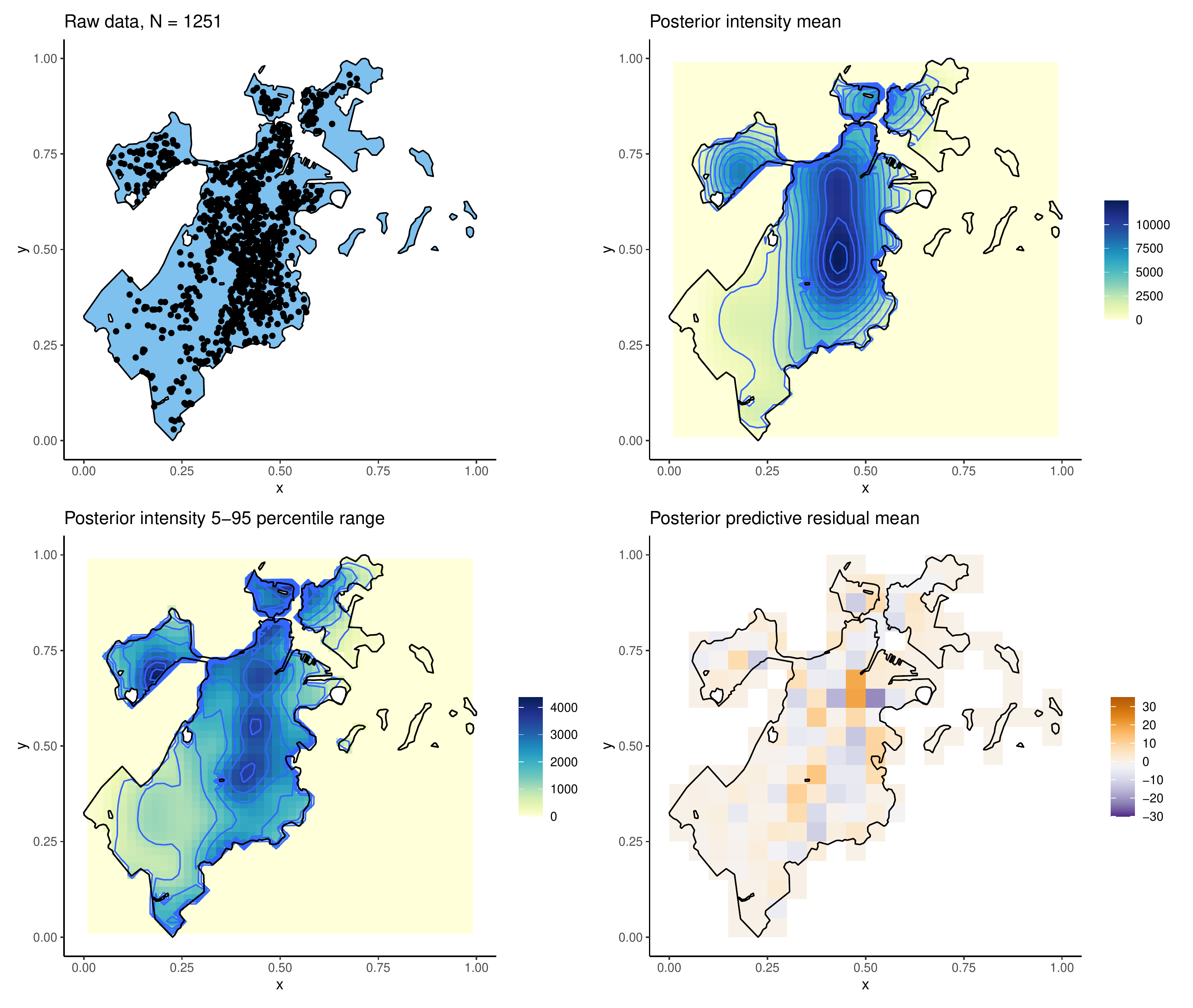}
\caption{Boston crime data: vandalism in the second quarter of 2017. The observed 
point pattern is shown in the top left panel. Under the density model, the top right
panel plots the posterior mean intensity estimate, and the bottom left panel the  
difference between the 95th and 5th percentile of the posterior distribution for the
intensity function. The bottom right panel plots the posterior mean estimates for
the predictive residuals.}
\label{fig:bostona}
\end{figure}

For an illustration with real data, we consider the point pattern of $n=1251$ locations 
in the city of Boston where vandalism occurred during the second quarter of year 2017;
see the top left panel of Fig. \ref{fig:bostona}.
In general, spatial point patterns of crime depict more clustering than what a 
NHPP can model. However, we use such data here to illustrate the spatial NHPP model 
over a non-trivial irregular domain, including model checking of the NHPP assumption.

The Boston City crime data and the Boston city boundary shape file in longitude 
and latitude format are publicly available online 
\citep[][]{crime-boston, boundary-boston}. We use the \texttt{R} \texttt{rmapshaper} 
package \citep[]{rmapshaper21} to smooth this complicated boundary while retaining its 
key spatial topology. The simplified boundary in the form of Multipolygons is then 
mapped to a subset of the unit square. To process the raw data, we remove entries 
with geo-location as NAs, project the vandalism incidence locations from longitude and 
latitude into Northing and Easting, and finally map the crime locations and city 
boundary points to the unit square.

We focus on inference results under the density model, implemented
with $C=0.01$, a $\text{Ga}(5, 0.1)$ prior for $\alpha$, and a truncated 
Poisson prior for $K$ with mean 20 and support on $[20, 60]$. 
Fig. \ref{fig:bostona} plots posterior mean and uncertainty estimates for the 
intensity of vandalism incidences. 
The posterior mean for the total intensity of vandalism in the second quarter of 2017 
is $1234$, with the 95\% posterior credible interval given by $(1167, 1303)$. 
The posterior distribution for $K$ has effective support on $[36, 52]$ and 
posterior mode at 40 with $\text{Pr}(K=40 \mid \text{data})=0.34$. 
%
%

For graphical model checking, we consider predictive residuals \citep{LG_2017}, 
defined as $N_{\text{pred}}(\mathcal{B}) - N_{\text{obs}}(\mathcal{B})$, where
$N_{\text{obs}}(\mathcal{B})$ and $N_{\text{pred}}(\mathcal{B})$ are respectively 
the observed and predicted number of points in $\mathcal{B}$, a subset of the 
spatial point process domain $\mathcal{D}$. To sample from the posterior distribution
of $N_{\text{pred}}(\mathcal{B})$, we draw from the 
$\text{Poisson}(\iint_{\mathcal{B}} \lambda_{\mathcal{D}}(x, y) \, \text{d}x \text{d}y)$
distribution for each posterior realization of $\lambda_{\mathcal{D}}(x, y)$.
%
%
%
In general, lack of fit may be due to the NHPP assumption for the point process 
that generates the particular point pattern and/or the model used for the NHPP 
intensity. A flexible prior probability model for the NHPP intensity is practically 
useful in that it allows focusing discrepancies in the residuals on the NHPP assumption.

To implement model checking with predictive residuals, we create a $20 \times 20$ grid 
over the unit square and select the subset of these $400$ square regions that overlap 
with the Boston city boundary $\mathcal{D}$ as the target regions to cover the entire 
Boston city. The bottom right panel of Fig. \ref{fig:bostona} plots the posterior mean 
estimates for the predictive residuals. The residuals in regions near the city 
boundary are evaluated based on only the subsets that overlap with $\mathcal{D}$. 
This residual analysis suggests a decent fit of the NHPP model. It is perhaps 
not surprising that the sub-regions with the more pronounced non-zero residual 
estimates correspond to parts of the city where the data suggest clustering,
for which a more general point process than the NHPP would be expected 
to provide better model fit.

%
%

\section{Discussion}
\label{discussion}

We have presented two models for spatial NHPP intensities over domains with irregular shapes. 
To our knowledge, this is the first treatment of this practically relevant problem with
methodology that supports general intensity function shapes and allows for full Bayesian 
inference, while avoiding any type of approximation of the NHPP likelihood.

In the more commonly studied setting of a regular domain, the two modelling approaches result 
in the same formulation for the NHPP intensity, which corresponds to a Bernstein-Dirichlet 
prior for the associated NHPP density. Hence, as a useful byproduct of the methodology,
we establish a connection between density and intensity estimation under Bernstein-Dirichlet 
priors. Relative to existing approaches that model directly the intensity function over 
regular domain, the proposed method arguably offers a substantially more practical 
inference framework. The prior model for the intensity function can be equivalently 
represented in terms of a prior for the total intensity over the observation domain 
and a prior for the density function. In contrast with related existing methods, 
the priors for the NHPP density and the total intensity are guaranteed to be compatible 
with the prior for the NHPP intensity.

The two proposed models for spatial NHPPs over irregular domains $\mathcal{D}$,
the intensity model (\ref{intensity-2d_irregular}) and density model (\ref{2d-density-ireg}), 
arise from different perspectives. The former model builds from truncating the 
Bernstein-Dirichlet density model over $\mathcal{D}$, whereas the latter constructs the 
irregular domain density as a mixture of truncated beta basis densities. 
The intensity model uses all $K^2$ basis densities $\{ \phi^{*}_{k_x, k_y} \}$
and relies on random weights, further adjusted by the normalizing constants $B_{k_x,k_y}$, 
to select appropriate basis members in constructing the intensity functional form. 
The density model is generally more efficient in the intensity representation, as it utilizes 
a subset of the $K^2$ basis densities $\{ \phi^{*}_{k_x, k_y} \}$, the size of such subset 
determined by the particular domain $\mathcal{D}$. For settings where a value for 
$K$ can be specified, possibly appealing to empirical experience with synthetic data examples, 
the intensity model offers the benefit of particularly simple and efficient model fitting. 
The density model affords more generality in the inference scheme by allowing uncertainty 
with respect to the number of basis densities, at the cost of a more involved posterior 
simulation method, which however does not require complex trans-dimensional computational 
techniques. For both models, the intensity representation through beta densities with 
specified parameters is essential for the practicality and computational efficiency 
of the inference methods for spatial NHPPs over irregular domains.

The proposed methods are also relevant for more general spatial point processes that build 
from the NHPP to allow hierarchically structured, clustering mechanisms. Current research 
is exploring Bayesian semiparametric spatial and space-time Hawkes processes for applications 
where it is important to address the spatial point process irregular domain.

\section*{Acknowledgment}
This research was supported in part by the National Science Foundation 
under award SES 1950902.

\bibliographystyle{biometrika}
\bibliography{bp_nhpp}

\newpage

\appendix

\section{Appendix: Prior expectation for the intensity function}

We use $\text{be}(\cdot\mid \cdot, \cdot)$ and $\text{ga}(\cdot \mid \cdot, \cdot)$ to denote the beta and gamma distributions' probability density functions. We use $\text{Be}(\cdot, \cdot)$ and $\text{Ga}(\cdot, \cdot)$ to denote the Beta and gamma distribution, $\mathbbm{1}_{\mathcal{D}}(\boldsymbol{s})$ to denote the indicator function of whether the point $\boldsymbol{s}$ in $\mathbb{R}$ or $\mathbb{R}^2$ is in the interval or region $\mathcal{D}$.
We write the spatial Bernstein basis functions over the unit square as $\phi_{k_x, k_y}(x, y) = \text{be}(x|k_x, K - k_x + 1)\text{be}(y|k_y, K-k_y +1)$, for $k_x, k_y = 1 \cdots, K$, and the spatial Bernstein basis functions over the irregular domain $\mathcal{D}$ as $\phi^*_{k_x, k_y}(x, y) = B_{k_x, k_y}^{-1}\text{be}(x|k_x, K - k_x + 1)\text{be}(y|k_y, K-k_y +1)$, 
where $B_{k_x, k_y}$ is the normalizing constant for density $\phi_{k_x, k_y}$ over the irregular domain $\mathcal{D}$. We use $\{S_{k_x, k_y}: k_x, k_y = 1\cdots K\}$ to denote the $K \times K$ unit square partitioning sets, $S_{k_x, k_y}^* $ the overlap between $S_{k_x, k_y}$ and $\mathcal{D}$, $S_{k_x, k_y}^*= S_{k_x, k_y} \cap \mathcal{D}$, and $J_K$ the index set for $S^*_{k_x, k_y}$ where $S^*_{k_x, k_y} \neq \emptyset$. Finally, we let 
$F_0(S_{k_x, k_y}^*)$ denote the probability of $S_{k_x, k_y}^*$ under distribution $F_0$.

Here, we derive the prior expectation of the intensity function under the 
different Poisson process models. 

For the temporal Poisson process model (2) developed in Section 2.1, we have:
\begin{eqnarray*}
\text{E}(\lambda(s)|\alpha, K) &=& \sum_{k = 1}^K 
\text{E}(V_k \mid \alpha) \, \text{be}(s|k, K-k +1) \\
&=& \frac{\alpha}{C}\sum_{k = 1}^K \frac{1}{K} \frac{K!s^{k-1}(1-s)^{K-k}}{(k-1)!(K-k)!} \\
&=& \frac{\alpha}{C} \sum_{m = 0}^{K-1} \frac{(K-1)!s^m (1-s)^{K-1-m}}{m!(K-1-m)!} \\
&=& \frac{\alpha}{C} \sum_{m=0}^{K-1}{K-1 \choose m} s^m (1-s)^{K-1-m}\\ 
&=& \frac{\alpha}{C}
\end{eqnarray*}
using the Binomial theorem. Note that the conditional prior expectation does not depend on $K$.
Finally, $\text{E}(\lambda(s)) =$ $\text{E}(\text{E}(\lambda(s) \mid \alpha)) =$
$\text{E}(\alpha)/C$.

For the spatial Poisson process model (5), over the unit square, we can write:
\begin{eqnarray*}
\text{E}(\lambda(x, y) \mid \alpha, K)  & =& \sum_{k_x = 1}^K \sum_{k_y = 1}^K 
\text{E}(V_{k_x, k_y} \mid \alpha) \, \text{be}(x|k_x, K-k_x + 1)\text{be}(y|k_y, K-k_y + 1) \\
&=& \frac{\alpha}{C} \, \frac{1}{K^{2}}
\sum_{k_x = 1}^K \sum_{k_y = 1}^K \text{be}(x|k_x, K-k_x + 1)\text{be}(y|k_y, K-k_y + 1) \\
& = & \frac{\alpha}{C}
\end{eqnarray*}
using the fact that $K^{-1} \sum_{m = 1}^K \text{be}(s|m, K - m + 1) = 1$, which is 
essentially a restatement of the Binomial theorem.

Similarly, for the irregular domain spatial model (7), developed in Section 3.1, we obtain:
\begin{align*}
\text{E}(\lambda_{\mathcal{D}}(x, y) \mid \alpha)
&= \sum_{k_x = 1}^K\sum_{k_y = 1}^K B_{k_x, k_y}\text{E}(V_{k_x, k_y} \mid \alpha)
\left(B_{k_x, k_y}^{-1}\text{be}(x|k_x, K-k_x + 1)\text{be}(y|k_y, K-k_y + 1)\right)  \\
&= \sum_{k_x = 1}^K\sum_{k_y = 1}^K \text{E}(V_{k_x, k_y} \mid \alpha)
\, \text{be}(x|k_x, K-k_x + 1)\text{be}(y|k_y, K-k_y + 1) \\ 
& = \frac{\alpha}{C} 
\end{align*}

Finally, for the irregular domain spatial model (10), developed in Section 3.2, we 
do not have an analytical expression for the prior mean intensity function. However,
we can obtain an approximate lower bound as follows:
\begin{align*}
    \text{E}(\lambda_{\mathcal{D}}(x, y) \mid \alpha,K) 
    &= \sum_{(k_x, k_y) \in J_K} \frac{\alpha F_0(S_{k_x, k_y}^*)}{C} \, \phi^*_{k_x, k_y}(x, y) \\
    &\approx \frac{\alpha}{C}\sum_{(k_x, k_y) \in J_K} \frac{1}{K^2} \, \phi^*_{k_x, k_y}(x, y)\\
    & \geq \frac{\alpha}{C}\sum_{(k_x, k_y) \in J_K} \frac{1}{K^2} \, B_{k_x, k_y}\phi^*_{k_x, k_y}(x, y)\\
    & \approx \frac{\alpha}{C} \sum_{k_x = 1}^K \sum_{k_y = 1}^K \frac{1}{K^2} \, B_{k_x, k_y}\phi^*_{k_x, k_y}(x, y) \\
    & = \frac{\alpha}{C} \sum_{k_x = 1}^K \sum_{k_y = 1}^K \frac{1}{K^2} \, \text{be}(x|k_x, K - k_x + 1) \text{be}(y|k_y, K - k_y + 1) = \frac{\alpha}{C}
\end{align*}
In step 2, we use the fact that $S_{k_x, k_y}^*$ is the overlap between the $K \times K$ unit square partition set $S_{k_x, k_y}$ and the irregular domain $\mathcal{D}$, and will have area either 
exactly equal to $1/K^{2}$, when $S_{k_x, k_y}^* =$ $S_{k_x, k_y}$, or area that can be approximated by $1/K^2$. In step 4, we use the fact that $B_{k_x, k_y} \approx 0$ for $(k_x, k_y) \notin J_K$.

\section{Appendix: Posterior simulation details}
\subsection{Temporal NHPP models}

Here we give posterior simulation details for model (2) in section 2.1. Given the number of basis $K$, the Markov Chain Monte Carlo algorithm consists of Gibbs or Metropolis update from the full conditionals for $\xi_i, V_k$ and $\alpha$. 
\begin{enumerate}
\item $\xi_i|-$
$$
    p(\xi_i = j|-) = \frac{V_j \text{be}(s|j, K-j + 1)}{\sum_{l = 1}^K V_{l}\text{be}(s|l, K-l+1)}
$$    
\item $V_k|-$ for $k = 1 \cdots K$, $M_k = \sum_{i = 1}^n \delta_{k}(\xi_i)$
$$
p(V_k|-) \propto \exp(-V_k)V_k^{M_k}V_k^{\alpha/K-1}\exp(-CV_k)
\propto \text{ga}(V_k|M_k + \alpha /K, C + 1)
$$
\item $\alpha|-$
\begin{align*}
    p(\alpha|-) &\propto \text{ga}(\alpha|a_\alpha, b_\alpha) \prod_{k= 1}^K \text{ga}(V_{k_x, k_y}|\alpha K^{-1}, C) \\
    &\propto \alpha^{a_\alpha-1}\exp(-b_\alpha \alpha)C^{\alpha}\Gamma(\alpha/K)^{-K}\prod_{k = 1}^K V_{k}^{\alpha/K}
\end{align*}
A metropolis step is implemented on the log scale with a normal random walk proposal density to sample from this full conditional. 

\end{enumerate}

The Markov Chain Monte Carlo algorithm for the density formulation of the temporal Poisson process under model (3) in section 2.1 consists of either a Metropolis or a Gibbs update from the following full-conditionals: \\
Let $k(s|\theta) = \sum_{k = 1}^K \mathbbm{1}_{\left((k-1)/K, k/K \right)}(\theta) \text{be}(s|k, K-k+1)$
\begin{enumerate}
    \item $\theta_i|\boldsymbol{\theta_{-i}},-$
\begin{equation*}
\begin{array}{c}
 p(\theta_i|\theta_{-i}, -) = \frac{\alpha q_0}{\alpha q_0 + H} \frac{k(s_i|\theta)f_0(\theta)}{q_0} + \frac{1}{\alpha q_0 + H} \sum_{j= 1}^{n^{*-}}k(s_i|\theta_j^{*-})n_j^- \delta_{\theta^{*-}_j}(\theta_i)\\
 q_0 = \int k(s_i|\theta)  f_0(\theta) \text{d}\theta = \sum_{j=1}^K\text{be}(s_i|j, K-j+1)\alpha/K \\
 H = \sum_{j=1}^{n^{*-}} k(s_i|\theta_j^{*-})n_j^{-}
\end{array}
\end{equation*}

where $ n^{*-}$ is the number of unique values, $\{\theta_j^{*-}: j=1 \cdots n^{*-}\}$ is the vector of unique values, and $\{n_j^{-}, j = 1 \cdots n^{*-}\}$ the vector of the number of observations that take value $\theta_j^{*-}$ in the vector $\boldsymbol{\theta_{-i}} = \{\theta_l: l \neq i\}$. 

\item $\Lambda|-$  
$$
\Lambda|- \sim \text{Ga}(\alpha + n, C + 1)
$$
where n is the number of points 

\item $\alpha|-$
\begin{align*}
p(\alpha|-) &\propto (\prod_{m = 1}^n(\alpha + m -1))^{-1}\alpha^{n^*} \text{ga}(\Lambda|\alpha, C)\text{ga}(\alpha|a_0, b_0)\\
& \propto (\prod_{m = 1}^n(\alpha + m -1))^{-1}\alpha^{n^*} \frac{C^\alpha}{\Gamma(\alpha)}\Lambda^{\alpha-1}\alpha^{a_0 -1} \exp(-b_0\alpha)
\end{align*}
A metropolis step is implemented with a normal proposal on the log scale. 

\item $K|-$
$$
p(K|-) \propto \prod_{i = 1}^n \left\{\sum_{k=1}^K \mathbbm{1}_{[\frac{k-1}{K}, \frac{k}{K}]}(\theta_i)\text{be}(s_i|k, K-k+1)\right\} \pi(K|\{K_{min}, \cdots, K_{max}\})
$$

The full conditional for $K$ is a discrete distribution and can be directly sampled from. 
\end{enumerate}

With each draw in the posterior sample for $\{(\theta_1 \cdots \theta_n), \alpha, K\}$, we sample $\{\omega_k: k = 1\cdots K\}$ from the following Dirichlet distribution

$$
\{\omega_k: k = 1\cdots K\} \sim \text{Dir}(\{\alpha/K +  \sum_{i = 1}^n \mathbbm{1}_{[\frac{k-1}{K}, \frac{k}{K}]}(\theta_i): k = 1\cdots K\})
$$

We obtain a draw from the posterior distribution of the intensity function $\lambda(s)$ and the density function $f(s)$ evaluated at location $s$ respectively, given $\{\omega_1, \cdots, \omega_K\}$ via the following functions 
$$
\begin{array}{c}
f(s) = \sum_{k =1}^K \omega_k \text{be}(s|k, K-k+1)\\
\lambda(s) = \Lambda \sum_{k =1}^K \omega_k \text{be}(s|k, K-k+1)
\end{array}
$$

\subsection{The intensity formulation for spatial NHPP over irregular domain}

The full hierarchical model for the intensity formulation over irreglar domain under (8) in section 3.1  
\begin{equation*}
\begin{array}{c}
\{(x_{i},y_{i}) \} \mid \bm{V}, \{ (\xi_{i},\eta_{i}) \}  \sim 
\left( \prod\nolimits_{k_x,k_y = 1}^{K} \exp( - V_{k_x, k_y} B_{k_x, k_y} ) \right) 
\prod_{i=1}^{n} \Lambda_\mathcal{D} \, \phi^{*}_{\xi_{i}, \eta_{i}}(x_{i},y_{i}) 
\notag \\
(\xi_{i},\eta_{i}) \mid \bm{V}  \stackrel{i.i.d.}{\sim}  
\sum_{k_x, k_y =1}^{K} \frac{V_{k_x, k_y} B_{k_x, k_y}} {\Lambda_\mathcal{D}} \,
\delta_{(k_x, k_y)}(\xi_{i},\eta_{i}),  \quad i=1,...,n  \\
\alpha, \bm{V}  \sim  \text{Ga}(\alpha \mid a_{\alpha},b_{\alpha})
\, \prod_{k_x, k_y = 1}^{K} \text{Ga}(V_{k_x, k_y} \mid \alpha K^{-2},C)
\end{array}
\end{equation*}
where $\bm{V}=$ $\{ V_{k_x, k_y}: k_x,k_y = 1,...,K \}$, $\Lambda_\mathcal{D} =$
$\sum\nolimits_{k_x,k_y = 1}^{K} V_{k_x, k_y} B_{k_x, k_y}$, and $
B_{k_x, k_y} = \int\!\int_D \text{be}(x|k_x, K-k_x+1)\text{be}(y|k_y, K-k_y+1) \text{d}x\text{d}y
$. $\{B_{k_x, k_y}, k_x, k_y =  1 \cdots K\}$ can be computed given $K$ and $\mathcal{D}$ before running the Markov Chain Monte Carlo algorithm to save computation time.

Given the number of basis $K$, the Markov Chain Monte Carlo algorithm consists of Gibbs or Metropolis update from the full conditionals for $\{\xi_i, \eta_i\}, V_{k_x, k_y}$ and $\alpha$:

\begin{enumerate}
    \item The full conditional for $\{\xi_i, \eta_i\}$ are discrete distributions 

$$
p(\xi_i = m, \eta_i = n|-) = \frac{V_{m, n} 
\text{be}(x_i|m, K-m +1)\text{be}(y_i|n, K-n+1)}{\sum_{p, q = 1}^KV_{p, q}\text{be}(x_i|p, K-p +1)\text{be}(y_i|q, K-q+1) } 
$$

\item The full conditional for $V_{k_x, k_y}, k_x, k_y = 1 \cdots K$ are independent Gamma distributions, which can be sampled directly from in a vectorized fashion.
Let $M_{k_x, k_y}$ be the number of latent variable pairs $(\xi_i, \tau_i)$ in step 1 that take value $(k_x, k_y)$. 

\begin{align*}
p(V_{k_x, k_y}| -) & \propto \exp(-V_{k_x, k_y}B_{k_x, k_y}) \prod_{i = 1}^n  \Lambda_\mathcal{D} \sum_{k_x, k_y = 1}^K \frac{V_{k_x, k_y}B_{k_x, k_y}}{\Lambda_{\mathcal{D}}} \delta_{(k_x, k_y)}(\xi_i, \eta_i) \phi^*_{k_x, k_y}(x_i, y_i)\\
& \quad \times \text{ga}(V_{k_x, k_y}|\alpha/K^2, C) \\ 
& \propto \exp(-V_{k_x, k_y}B_{k_x, k_y}) \prod_{i = 1}^n (V_{k_x, k_y}B_{k_x, k_y})^{\delta_{(k_x, k_y)}(\xi_i, \eta_i)} \text{ga}(V_{k_x, k_y}|\alpha/K^2, C) \\
& \propto \text{ga}(V_{k_x, k_y}|M_{k_x, k_y} + \alpha/K^2, C+B_{k_x, k_y})
\end{align*}

\item The full conditional for $\alpha$ is 
\begin{align*}
    p(\alpha|-) &\propto \text{ga}(\alpha|a_\alpha, b_\alpha) \prod_{k_x, k_y = 1}^K\text{ga}(V_{k_x, k_y}|\alpha K^{-2}, C) \\
    &\propto \alpha^{a_\alpha-1}\exp(-b_\alpha \alpha)C^{\alpha}\Gamma(\alpha/K^2)^{-K^2}\prod_{k_x, k_y = 1}^K V_{k_x, k_y}^{\alpha K^{-2}}
\end{align*}
A metropolis step is implemented on the log scale with a normal random walk proposal density to sample from this full conditional. 
\end{enumerate}

\subsection{The density formulation for spatial NHPP over irregular domain}

The full hierarchical model for the density formulation over irregular domain under model (11) in section 3.2 is 

\begin{equation*}
\begin{array}{c}
\{ (x_{i},y_{i}) \} \mid \{\bm{z}_i\}, \Lambda_\mathcal{D}, K  \sim 
\exp( - \Lambda_\mathcal{D} ) 
\prod_{i=1}^{n} \Lambda_\mathcal{D} \, \sum_{k_x, k_y \in J_k} \mathbbm{1}_{S^*_{k_x, k_y}}(\bm{z_i})\phi^{*}_{k_x, k_y}(x_{i},y_{i}) 
\\
(x_i, y_i), \bm{z_i} \in \mathcal{D}, \,\,\,\, \bm{z_i}\mid F \stackrel{i.i.d}{\sim} F, 
\,\,\, i = 1 \cdots, n  \\
F\mid \alpha \sim \text{DP}(\alpha, F_0) \,\,\, F_0(\cdot) \equiv \text{Unif}(\mathcal{D}) \\ 
\Lambda_\mathcal{D}\mid \alpha \sim \text{Ga}(\alpha, C) \,\,\, \alpha \sim \text{Ga}(\alpha \mid a_{\alpha},b_{\alpha}) \,\,\, K \sim \pi(K \mid \{K_{\min},\cdots, K_{\max}\})
\end{array}
\end{equation*}

The Markov Chain Monte Carlo algorithm consists of either Metropolis or Gibbs update from the following full-conditionals: 
\begin{enumerate}

\item $\boldsymbol{z_i}|\boldsymbol{z_{-i}}, \alpha, K$, where $\boldsymbol{z_i}$ is the bivariate continuous latent variable. 

Let $k^*(\boldsymbol{s_i}|\boldsymbol{z_i}) = \sum_{(k_x, k_y)\in J_k} \mathbbm{1}_{S^*_{k_x,k_y}}(\boldsymbol{z_i})W^*_{k_x, k_y, i}$, where $W^*_{k_x, k_y, i} = \phi^*_{k_x, k_y}(x_i, y_i)$ is a constant. 

\begin{equation*}
\begin{array}{c}
p(\boldsymbol{z_i}|\boldsymbol{z_{-i}}, \boldsymbol{s_i}) = \frac{\alpha q_0}{\alpha q_0 + H} \frac{k^*(\boldsymbol{s_i}|\boldsymbol{z})f_0(\boldsymbol{z})}{q_0} + \frac{1}{\alpha q_0 + H} \sum_{j=1}^{n^{*-}} k^*(\boldsymbol{s_i}|\boldsymbol{z_j^{*-}}) n_j^- \delta_{\boldsymbol{z_j^{*-}}}(\boldsymbol{z_i}) \\
= \frac{\alpha q_0}{\alpha q_0 + H} q(\boldsymbol{z}|\boldsymbol{s_i}) + \frac{1}{\alpha q_0 + H} \sum_{j=1}^{n^{*-}} k^*(\boldsymbol{s_i}|\boldsymbol{z_j^{*-}}) n_j^- \delta_{\boldsymbol{z_j^{*-}}}(\boldsymbol{z_i}) \\
q_0 = \int k^*(\boldsymbol{s_i}|\boldsymbol{z})f_0(\boldsymbol{z})\text{d}\boldsymbol{z} = \sum_{(k_x, k_y)\in J_k} W^*_{k_x, k_y, i} \cdot \mid S^*_{k_x, k_y} \mid / \mid\mathcal{D}\mid \\
q(\boldsymbol{z}|\boldsymbol{s_i})  = \sum_{(k_x, k_y)\in J_k} W^*_{k_x, k_y, i}q_0^{-1} \mathbbm{1}_{S^*_{k_x, k_y}}(\boldsymbol{z})\mid \mathcal{D}\mid^{-1} \\
= \sum_{(k_x, k_y)\in J_k} \frac{W^*_{k_x, k_y, i} \mid S^*_{k_x, k_y}\mid}{\sum_{m, n}W^*_{m, n, i} \mid S^*_{k_x, k_y}\mid}\mathbbm{1}_{S^*_{k_x, k_y}}(\boldsymbol{z})\mid S^*_{k_x, k_y}\mid^{-1}\\
H = \sum_{j = 1}^{n^{*-}} k^*(\boldsymbol{s_i}|z_j^{*-}) n_j^{-}
\end{array}
\end{equation*}
where $ n^{*-}$ is the number of unique values,  $\{\boldsymbol{z_j^{*-}}: j=1 \cdots n^{*-}\}$ is the vector of unique values, and $\{n_j^{-}, j = 1 \cdots n^{*-}\}$ is the vector of the number of observations that take value $z_j^{*-}$ in the vector $\boldsymbol{z_{-i}} = \{\boldsymbol{z_l}: l \neq i\}$. 
\item $\Lambda_\mathcal{D}|-$
$$
\Lambda_\mathcal{D}|- \sim \text{Ga}(\alpha + n, C + 1)
$$
where $n$ is number of points.
\item $\alpha|-$
\begin{align*}
p(\alpha|-) &\propto (\prod_{m = 1}^n(\alpha + m -1))^{-1}\alpha^{n^*} \text{ga}(\Lambda_\mathcal{D}|\alpha, C)\text{ga}(\alpha|a_0, b_0)\\
& \propto (\prod_{m = 1}^n(\alpha + m -1))^{-1}\alpha^{n^*} \frac{C^\alpha}{\Gamma(\alpha)}\Lambda_\mathcal{D}^{\alpha-1}\alpha^{a_0 -1} \exp(-b_0\alpha)
\end{align*}
A metropolis step is implemented with a normal proposal on the log scale. 

\item $K|-$ 
$$
p(K|-) \propto \prod_{i = 1}^n \left\{\sum_{k_x, k_y \in J_K} \mathbbm{1}_{S^*_{k_x, k_y}}(\boldsymbol{z_i})W^*_{k_x, k_y, i}\right\} \pi(K|\{K_{min}, \cdots, K_{max}\})
$$
The full conditional for K is a discrete distribution and can be directly sample from. 
\end{enumerate}

With each draw in the posterior sample for $\{\Lambda_\mathcal{D}, \{\bm{z_i}\}, \alpha, K\}$, we obtain a draw from the posterior distribution of $\{\omega^*_{k_x, k_y}:(k_x, k_y) \in J_K\}$ by sampling from the following Dirichlet distribution:
$$  
\{\omega^*_{k_x, k_y}: k_x, k_y = 1\cdots K \} \sim \text{Dir}(\{\alpha/\mid S^*_{k_x,k_y}\mid + \sum_{i = 1}^n \mathbbm{1}_{S^*_{k_x, k_y}}(\boldsymbol{z_i}): k_x, k_y = 1\cdots K\})
$$
We obtain a draw from the posterior distribution of the intensity function $\lambda_\mathcal{D}(\bm{s})$ and the density function $f_\mathcal{D}(\bm{s})$ evaluated at location $\bm{s}=(x, y)$ respectively, given $\{\omega^*_{k_x, k_y}: k_x, k_y = 1\cdots K\}$, via the following function
$$
\begin{array}{c}
f_\mathcal{D}(x, y) = \sum_{(k_x, k_y)\in J_K} \omega_{k_x, k_y}^*\phi^*_{k_x, k_y}(x, y)\\
\lambda_\mathcal{D}(x, y) = \Lambda_\mathcal{D} \sum_{(k_x, k_y)\in J_K} \omega_{k_x, k_y}^*\phi^*_{k_x, k_y}(x, y)
\end{array}
$$

\end{document}